\documentclass[10pt]{article}
\usepackage[latin1]{inputenc}
\usepackage{amsmath}
\usepackage{amsfonts}
\usepackage{amssymb}
\usepackage{hyperref}
\usepackage{graphicx}
\usepackage[margin=2cm]{geometry}

\title{Disentangling multi-level systems: averaging, correlations and memory}

\author{Jeroen Wouters, Valerio Lucarini \vspace{0.5cm}\\
Meteorologisches Institut,\\
University of Hamburg,\\
KlimaCampus\\
Grindelberg 7, 20144 Hamburg, Germany
}

\begin{document}
\maketitle

\abstract{We consider two weakly coupled systems and adopt a perturbative approach based on the Ruelle response theory to study their interaction. We propose a systematic way to parametrize the effect of the coupling as a function of only the variables of a system of interest. Our focus is on describing the impacts of the coupling on the long-term statistics rather than on the finite-time behaviour. By direct calculation, we find that, at first order, the coupling can be surrogated by adding a deterministic perturbation to the autonomous dynamics of the system of interest. At second order, there are additionally two separate and very different contributions. One is a term taking into account the second order contributions of the fluctuations in the coupling, which can be parametrized as a stochastic forcing with given spectral properties. The other one is a memory term, coupling the system of interest to its previous history, through the correlations of the  second system. If these correlations are known, this effect can be implemented as a perturbation with memory on the single system. In order to treat this case, we present an extension to Ruelle's response theory able to deal with integral operators. We discuss our results in the context of other methods previously proposed to disentangle the dynamics of two coupled systems. We emphasize that our results do not rely on assuming a time scale separation, and, if such a separation exist, can be used equally well to study the statistics of the slow as well as that of the fast variables. By recursively applying the technique proposed here, we can treat the general case of multilevel systems.}

\section{Introduction}

The dynamics of complex systems can often be partitioned in such a way to highlight a multi-level structure, where each level consists of a subset of  variables that interact strongly with each other and are weakly coupled with variables belonging to other levels of the system. Often, the dynamics of each level takes place in distinct spatial and temporal scales, so that it is possible to hierarchically arrange the variables from the most to the least relevant, in view of the properties of interest. In practical applications, the observation or the simulation at high resolution of processes occurring on small temporal and spatial scales is very challenging. It is crucial to devise strategies for accounting, at least in an approximate way, the impact of the fast, almost inaccessible variables, onto the slow variables, which are more usually the main object of investigation. In the language of dynamical systems theory, the overarching goal is to define rigorously or empirically an effective dynamics for the slow manifold of the system able to take into account the influence of the, enslaved, fast manifold. Otherwise, the presence of such very different scales make these systems hard to simulate, so that they are often referred to as stiff problems \cite{fatkullin_computational_2004}, and hard to observe in practical terms, because no overall optimal spatial and temporal resolution for the collection of data can be defined, with the result that some variables are poorly described, whereas other variables have redundant observations.  

In these terms, the climate system provides an especially interesting case study. In the study of geophysical fluid dynamics, one often encounters a wide range of scales. The spatial scales involved in atmospheric dynamics can range from $10^7 \text{m}$ down to $10^{-6} \text{m}$. Temporal scales can vary from $10^{16} s$ to $10^{-6} s$ \cite{lucarini_modelling_2011,vallis_atmospheric_2006}. Here, one has to take into account coupling between different subsystems, which feature dynamics taking place in vastly different ranges, as clarified when considering separately the atmosphere, the oceans, the biosphere, and the criosphere \cite{Peixoto,saltzman_dynamical_2001}. We still do not have a standard formulation of a climate model, because, depending on the class of problems we want to tackle, we need to resort to different approximations and even different formulations of the relevant dynamics. The state-of-the-art climate models used for the IPCC projections \cite{ipcc} cannot be used to study the last million years of Earth's history, not only because they are too computationally expensive, but because they do not take into the right considerations the processes which dominate the Earth dynamics on such long time scales. On the other hand, Earth Models of Intermediate complexity \cite{plasim,rahmstorf}, which are tailored for such problems, are definitely not competitive if our aim is to describe the variability of the present climate and the short-term climate change. The problem of providing a satisfactory description of the dynamical processes that cannot be explicitly be represented has long been addressed at various levels of rigour by the climate community \cite{orrell_model_2003,wilks_effects_2005,imkeller_stochastic_2001,
hasselmann_stochastic_1976,arnold_hasselmanns_2001,saltzman_dynamical_2001,palmer_stochastic_2009}. The solutions that are being used in practical applications today go by the name of parametrizations. Processes that are not explicitly described are approximated via functions (deterministic or, more recently, stochastic) of the variables of the model, basically along the lines of classical closure theories in fluid dynamics for the Reynolds' stress tensor. Given the approximate and empirical nature of the parametrizations, a problem that often arises is their robustness with respect to changes in the considered modelling set-up, such as the case when additional variables are included in the model or different large scale conditions are considered. Retuning the parametrization is definitely a delicate aspect of climate modelling and a major hurdle towards the goal of obtaining flexible climate models able to obtain the so-called seamless simulation of its statistical properties and seamless (probabilistic) prediction of its state. 

In this paper, we tackle this class of problems from a rather general point of view. We will consider a two-level system, consisting of two weakly coupled systems $X$ and $Y$. We focus our attention on the $X$ system. We want to derive \textit{ab initio} a suitable parametrization of its coupling with the $Y$ system. 

For this purpose, various reduction techniques have been proposed in the literature. A classic approach is the averaging method \cite{arnold_hasselmanns_2001, kifer_recent_2004}, which assumes that there is a vast time-scale separation between the two systems $X$ and $Y$. Such method allows to derive a renormalised dynamics for the $X$ system whose trajectories converge on finite time scales to those of the original dynamics. An interesting use of the averaging method for deriving a simplified dynamics has been proposed in ~\cite{abramov_suppression_2011}, where a sort of mean field ansatz is taken. In statical mechanics, projector operator techniques which allow to formally eliminate certain variables have found a great deal of success, as in the case of the Mori-Zwanzig equation \cite{zwanzig_nonequilibrium_2001}.

In the present manuscript, we take a different route, based on the response theory developed by Ruelle\cite{ruelle_differentiation_1997,ruelle_review_2009,lucarini08}. Such a theory provides explicit formulas for computing how the long-time averages of Axiom A dynamical systems (described by the so called SRB measure~\cite{young_what_2002}) change as a result of small perturbations to the flow. This approach is especially useful for studying the impact of changes in the internal parameters of a system or of small modulations to the external forcing \cite{reick02,cessac07,majda07,lucarini09,lucarini_statistical_2011}. According to the theory, the changes in the observables can be expressed basically as a power series in orders of the strength of the forcing, so that one should assume, in our case, that the coupling between the $X$ and $Y$ systems - which acts as perturbation to the dynamics of $X$ - should be small. We derive explicit expressions for the perturbations in the statistics of the $X$ system due to the coupling at the first and second order. We explain how these can be parametrized in terms of a surrogate deterministic forcing, multiplicative noise, and memory term, so that we construct an approximated projected dynamics on the $X$ variables. Our approach does not make use of any large time scale separation between the $X$ and the $Y$ variables, and does not even assume that the $X$ variables are slower than the $Y$ variables. 

This article is structured as follows. In Section~\ref{sec:response_theory} we give a short review of Ruelle's response theory at higher orders. In Section~\ref{sec:multilevel} we then use this theory to derive the response of the system $X$ to the coupling with the system $Y$. We derive the first and second order terms and present an expression for a surrogate dynamics of $X$ alone. The correction to the unperturbed dynamics will include a deterministic forcing, a stochastic forcing, and a forcing given by a term which takes into account the history of the system. We present also a diagrammatic representation of the most relevant contributions. In Section \ref{comparison} we briefly discuss our results in comparison to classic variables reduction methods such as the averaging method \cite{arnold_hasselmanns_2001, kifer_recent_2004} and the projection operator method \cite{zwanzig_nonequilibrium_2001}. In Section \ref{sec:conclusions} we  present our conclusions and perspectives for future work. In Appendix~\ref{sec:memory_response}, we describe how to extend the Ruelle response formula to forcings describing memory effects, which we use to treat the memory terms resulting at the second order of perturbation. 

\section{Response theory}
\label{sec:response_theory}
Ruelle~\cite{ruelle_nonequilibrium_1998} recently derived explicit formulas for describing the smooth dependence of the SRB measure of Axiom A dynamical systems to small perturbations of the flow. Such response theory boils down to a Kubo-like perturbative expression connecting the terms describing the linear and nonlinear response of the system as expectation values of observables on the unperturbed measure. At order $n$ of nonlinearity, such expectation values can be written as $n-$uple convolution of a causal Green function with the time-delayed perturbative fields, so that at every order Kramers-Kronig relations can be written for the Fourier transform of the Green function, the so-called susceptibility ~\cite{lucarini08}. In order to clarify the goal of response theory and set our notation, we present a simple derivation of the linear and higher-order nonlinear response formulas obtained by introducing the so-called Dyson interaction picture for the operators. Given a continuous time dynamical system $x_t \in M$ with dynamics
\begin{align}
\dot{x_t} = F(x_t)
\end{align}
and a corresponding flow function $f^t$ relating $x_t$ to the initial conditions $x_0$, $x_t=f^t(x_0)$, we assume that an invariant measure $\rho_0$ exists \cite{young_what_2002}, describing long-term averaged quantities:
\begin{align}
\rho_0(A)=\lim_{T \rightarrow \infty} \frac{1}{T} \int_0^T d \tau \int l(dx) A(f^\tau (x)) \,.
\end{align}
We want to know how such averages, given by the so called Sinai-Ruelle-Bowen (SRB) invariant measure $\rho_0$, are affected by small changes in the dynamics of the system. Let's assume that the dynamics of the system is altered by adding a time independent perturbation $\Psi(x)$:
\begin{align}
\dot{x_t} = F(x_t) + \Psi(x_t) \,,
\end{align}
with $\tilde{f}^t$ the perturbed flow function, so that we end up with a modified SRB measure $\tilde{\rho}$:
\begin{align}
\tilde{\rho}(A)=\lim_{T \rightarrow \infty} \frac{1}{T} \int_0^T d \tau \int l(dx) A(\tilde{f}^\tau (x)) \,.
\end{align}
If the strength of the perturbation is small, the corresponding change in the SRB measure of the system will also in general be small. The occurence of resonances has been studied in~\cite{ruelle_review_2009,cessac_does_2007,cessac_linear_2007}, but this is not expected to occur in system of physical interest (see \cite{colangeli_nonequilibrium_2010} and references therein). Hence, we can expand the perturbed SRB measure in powers of the perturbation $\Psi$. In order to do so, we first expand the evolved observable $A(\tilde{f}_t (x))$. This amounts to constructing a Dyson series, where the perturbation to the dynamical system plays the role of the interaction Hamiltonian. Using the chain rule, the time-dependence of an observable $A$ at time $t$ is described by
\begin{align}
\frac{dA(\tilde{f}^t(x_0))}{dt} = \mathcal{L} A(\tilde{f}^t(x_0)) = (\mathcal{L}_0 + \mathcal{L}_1) A(\tilde{f}^t(x_0)) \,,
\end{align}
where 
\begin{align}
\mathcal{L}_0 A(x) = F(x)\cdot\nabla_x A(x) \,, \qquad \mathcal{L}_1 = \Psi(x)\cdot\nabla_x A(x) \,, \qquad \mathcal{L} = \mathcal{L}_0 + \mathcal{L}_1.
\label{eq:propagators}
\end{align}
By defining an \textit{interaction} observable as $A_I(\tau,x_0) = \Pi_0(-\tau) A(\tilde{f}^\tau (x_0)) = \exp(-\mathcal{L}_0 \tau) \exp(\mathcal{L} \tau) A(x_0)$ with $\Pi_0(\tau)=\exp(\mathcal{L}_0 \tau)$, we can make an expansion of $A$ in orders of the perturbation $\Psi$. This observable $A_I$ obeys the differential equation
\begin{align}
\frac{d A_I(\tau,x_0)}{d\tau} = \mathcal{L}_I(\tau) A_I(\tau,x_0) \,, \label{eq:interaction}
\end{align}
where
\begin{align}
\mathcal{L}_I (\tau) = \Pi_0(-\tau) \mathcal{L}_1 \Pi_0( \tau) \,.
\end{align}
Note that the operator $\mathcal{L}_I$ is of order $\Psi$. A formal solution to Eq.~\ref{eq:interaction} is given by Duhamel's formula:
\begin{align}
A_I(\tau,x_0) = A_I(0,x_0) + \int_{0}^\tau d s_1 \mathcal{L}_I(s_1) A_I(s_1,x_0) \,.
\end{align}
By iterating this solution and making use of the fact that $A_I(\tau,x_0)=\exp(-\mathcal{L}_0 \tau) A(\tilde{f}^\tau (x_0))$, we get an expression for the evolved observable $A(\tilde{f}^\tau (x_0))$ in terms of $A(0,x_0)$:
\begin{align}
A(\tilde{f}^\tau (x_0)) = & \Pi_0(\tau) A(x_0) + \int_{0}^\tau ds_1 \Pi_0(\tau-s_1) \mathcal{L}_1 \Pi_0(s_1) A(x_0) \nonumber \\
& + \int_{0}^\tau d s_1 \int_{0}^{s_1} d s_2 \Pi_0(\tau-s_1) \mathcal{L}_1 \Pi_0(s_1-s_2) \mathcal{L}_1 \Pi_0(s_2) A(x_0) \nonumber +O(\Psi^3) \,.
\end{align}
Such an expression for the expectation values under perturbed dynamics $A(\tilde{f}^\tau (x_0))$ in terms of the unperturbed dynamics $\Pi_0(\tau) A(x_0) = A(f^\tau (x_0))$ allows us to expand the SRB measure $\tilde{\rho}$ around $\rho_0$. The perturbed SRB measure $\tilde{\rho}$ can be written as
\begin{align}
\tilde{\rho}(A) &= \lim_{T \rightarrow \infty} \frac{1}{T} \int_0^T d\tau \int l(dx) \left( \Pi_0(\tau) A(x) + \int_{0}^\tau ds_1 \Pi_0(\tau-s_1) \mathcal{L}_1 \Pi_0 (s_1) A(x) + O(\Psi^2) \right) \nonumber  \\
&=\rho_0(A)+\delta\rho^{(1)}(A)+O(\Psi^2) \,,
\end{align}
where the first term in this sum gives the unperturbed measure $\rho_0$, while the second term is the linear response to $\Psi$. We can rewrite it as
\begin{align}
\delta \rho^{(1)} (A) &= \lim_{T \rightarrow \infty} \frac{1}{T} \int_0^T d s_1 \int_{s_1}^{T} d \tau \int l(dx) \Pi_0(\tau-s_1) \mathcal{L}_1 \Pi_0(s_1) A(x) \nonumber \\
&= \lim_{T \rightarrow \infty} \frac{1}{T} \int_0^T d s_1 \int_0^{T-s_1} d u \int l(dx) \Pi_0(u) \mathcal{L}_1 \Pi_0(s_1) A(x) \nonumber \\
&= \int_0^\infty ds_1\rho_0(\mathcal{L}_1 \Pi_0(s_1) A)=\int_0^\infty ds_1\int \rho_0(dx)\Psi(x)\cdot\nabla_x A(f^{s_1}(x)))\nonumber \\&=\int_0^\infty ds_1G^{(1)}_{A,\Psi}(s_1)\label{drho1} \,.
\end{align}
This shows that the linear response can be written in terms of averages taken in the unperturbed system. Carrying on the calculation, we obtain the following expression for the higher order terms:
\begin{align}
\tilde{\rho} ( A) &=  \rho_0 (A) + \int_0^\infty ds_1 \rho_0 (\mathcal{L}_1 \Pi_0(s_1) A ) + \int_0^\infty d s_1 \int_0^{s_1} d s_2 \rho_0 ( \mathcal{L}_1 \Pi_0(s_1) \mathcal{L}_1 \Pi_0(s_2) A ) +O(\Psi^3)\nonumber \\
&=\rho_0(A)+\delta\rho^{(1)}(A)+\delta\rho^{(2)}(A)+O(\Psi^3) \,,
\label{eq:srb_perturbation}
\end{align}
where
\begin{align}
\delta \rho^{(2)} (A) &= \int_0^\infty d s_1 \int_0^{s_1} d s_2 \rho_0( \mathcal{L}_1 \Pi_0(s_1) \mathcal{L}_1 \Pi_0(s_2) A )\nonumber \\ &=\int_0^\infty ds_1\int_0^\infty ds_2\int \rho_0(dx)\Psi(x)\cdot\nabla_{x} \Psi(f^{s_1}(x))\cdot\nabla_{f^{s_1}(x)} A(f^{s_1+s_2}(x)))\nonumber \\&=\int_0^\infty ds_1\int_0^\infty ds_2G^{(2)}_{A,\Psi}(s_1,s_2)\label{drho2} \,.
\end{align}

\section{Two-level systems}
\label{sec:multilevel}
We will now make use of the response formulae derived in Section~\ref{sec:response_theory} to study how the statistical properties of a dynamical system which is composed, in the unperturbed case, by two independent subsystems, change when a weak coupling is introduced. Let's assume that the variables $x$ of the dynamical system can be separated into two groups of variables $X$ and $Y$
\[ x = \begin{pmatrix}
X \\ Y
\end{pmatrix} \,,
\]
so that the unperturbed dynamical system can be considered as given by the uncoupled dynamics:
\begin{align}
\dot{X} &= F_X (X) \,, \nonumber \\
\dot{Y} &= F_Y (Y) \,.
\label{eq:uncoupled_dyn}
\end{align}
A weak coupling is then introduced between the two subsystems as follows:
\begin{align}
\dot{X} &= F_X (X) + \Psi_X(X,Y) \,, \nonumber \\
\dot{Y} &= F_Y (Y) + \Psi_Y(X,Y) \,. \label{eq:coupled_dyn}
\end{align}
The unperturbed and perturbed evolution operators $\mathcal{L}_0$ and $\mathcal{L}_1$ are now given by
\begin{align}
\mathcal{L}_0 = F_X(X)\cdot\nabla_X + F_Y(Y)\cdot\nabla_Y\label{l0}
\end{align}
and
\begin{align}
\mathcal{L}_1 &= \Psi_X(X,Y)\cdot\nabla_X + \Psi_Y(X,Y)\cdot \nabla_Y \nonumber \\
&= \Psi^T \cdot \nabla \label{l1} \,,
\end{align}
where $\Psi^T = \begin{pmatrix} \Psi_X & \Psi_Y \end{pmatrix}$ and $ \nabla^T = \begin{pmatrix} \nabla_X & \nabla_Y \end{pmatrix}$.

\subsection{First order}
We derive from Eqs.~\ref{eq:propagators} and~\ref{eq:srb_perturbation} that the first order response to $\mathcal{L}_1$ is given by
\begin{align}
\delta_\Psi^{(1)} \rho (A) &= \int_0^\infty d s_1 \rho_0(\mathcal{L}_1 \Pi_0(s_1) A) \\
&= \int_0^\infty d s_1 \int \rho_0(dx) \Psi^T(x) \nabla (A \circ f^{s_1}) (x) \,,
\end{align}
where $\rho_0$ is the SRB measure corresponding to the uncoupled dynamics of Eq.~\ref{eq:uncoupled_dyn}. The response is an integral over time of the response function $G_{A_\Psi}^{(1)}(\tau)$ describing the delayed impact of a forcing in the past over a time span of $\tau$:
\begin{align}
\delta_\Psi^{(1)} \rho (A) =& \int_0^\infty d s_1 G_{A_\Psi}^{(1)}(s_1) \nonumber \\
G_{A,\Psi}^{(1)}(s_1) =& \int \rho_0 (dx) \Psi^T(x)\cdot \nabla ( A \circ f^{s_1}) (x)\label{dpsi1} \,.
\end{align}
By exploiting the fact that the unperturbed dynamics of the $X$ and $Y$ subsystems are uncoupled, this becomes
\begin{align}
G_{A,\Psi}^{(1)} (\tau) = \int \rho_0 (dx) \Psi^T(x)\cdot\nabla A \left( \begin{pmatrix}
f_X^\tau (X) \\ f_Y^\tau (Y) 
\end{pmatrix}
 \right) \,,
\end{align}
where $f^\tau_X$ and $f^\tau_Y$ represent the unperturbed flows of the $X$ and $Y$ subsystems, respectively. We are only interested in the evolution of observables on the $X$ system so we  consider observables $A$ that depend on $X$ variables only:
\begin{align}
A \left(
\begin{pmatrix}
X\\ Y 
\end{pmatrix}\right) = A(X) \,.
\end{align}
Therefore, the response function $G_{A,\Psi}^{(1)}$ can be written as
\begin{align}
G_{A,\Psi}^{(1)} (\tau) = \int \rho_0 \begin{pmatrix}
dX\\ dY 
\end{pmatrix}\Psi_X (X, Y ) \cdot \nabla_{X} A(f_X^\tau (X)) \,. \label{meas}
\end{align}
Since the dynamics of the unperturbed system is decoupled, its SRB measure $\rho_0$ can be written as a product of the SRB measure $\rho_{0,X}(dX)$ and $\rho_{0,Y}(dY)$. We have:
\begin{align}
G_{A,\Psi}^{(1)} (\tau) = \int \rho_{0,X} (dX) \langle \Psi_{X}(X,Y) \rangle_{\rho_{0,Y}} \cdot \nabla_{X} A( f_X^\tau (X) ) \,.
\label{respo}
\end{align}
Such a first order perturbation can be represented in a diagrammatic way as depicted in Figure~\ref{fig:1st_order}, where the arrow indicates the direction of the interaction and the time variable refers to the presence of a time-delayed, causal effect on the system of interest of the external perturbation.  

Equation \ref{respo} can be used to derive a parametrization of the first order effect of coupling the $X$ and $Y$ subsystems on the $A(X)$ observables, expressed as a function of the $X$ variables only. In other terms, we want to derive an expression for a perturbation vector field $M_1$ such that if we consider the dynamical system
\begin{align}
\frac{d X}{dt} = F_X(X) + M_1(X) \,, \label{eq:first_model}
\end{align}
where $F_X(X)$ is the same as in Eq. \ref{eq:uncoupled_dyn}, up to the first order the expectation values of the $A$ observables over the SRB invariant measure $\hat{\rho}_1$ of the dynamical system given in Eq. \ref{eq:first_model} and over the invariant measure $\tilde{\rho}_{X,Y}$ are identical. The invariant measure $\rho_{0} (dX)$ for the unperturbed $M_1=0$ system is, obviously, identical to partial measure $\rho_{0,X} (dX)$ introduced in Eq. \ref{meas}. Therefore, we must impose that the first order effect in Eq. \ref{eq:first_model} due to $M_1$ is the same as the first order correction due to the coupling between the $X$ and the $Y$ variables introduced in Eq.  \ref{eq:coupled_dyn}.  We discover that by choosing:  
\begin{align}
M_1(X) := \langle \Psi_{X}(X,Y) \rangle_{\rho_{0,Y}} \label{eq:subst_const}
\end{align}
and plugging this expression in Eq. \ref{drho1}, we obtain that up to the first order in $\Psi$ and $\forall A$:
\begin{align}
\delta\rho_{M_1}^{(1)}(A)=\delta_\Psi^{(1)}\rho(A) \,,
\end{align}
such that
\begin{align}
\tilde{\rho}_{X,Y}(A) =\rho_{0,X}(A)+\delta_\Psi^{(1)}\rho(A)+O(\Psi^2)= \hat{\rho}_1(A)+O(\Psi^2)=\hat{\rho}_0 (A) +\delta\rho_{M_1}^{(1)}(A)+ O(\Psi^2) \,.\label{o1}
\end{align}
Concluding, the effect on $X$ of its coupling with $Y$ can be described at first order with a mean-field theory constructed by taking the $Y$-ensemble average of the coupling, where the averaging is taken on the uncoupled $Y$ dynamics. 

\subsection{Second order}
\begin{figure}
\centering
\includegraphics[scale=0.5]{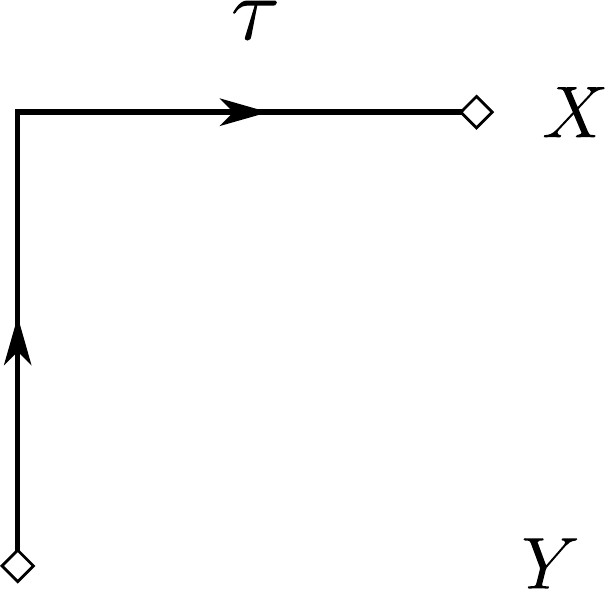}
\caption{First order response: diagram of the term describing the impact of the averages of the $Y$ variables $G^{(1)}_{A,\Psi} (\tau)$.}
\label{fig:1st_order}
\end{figure}

\begin{figure}
\centering
\includegraphics[scale=0.5]{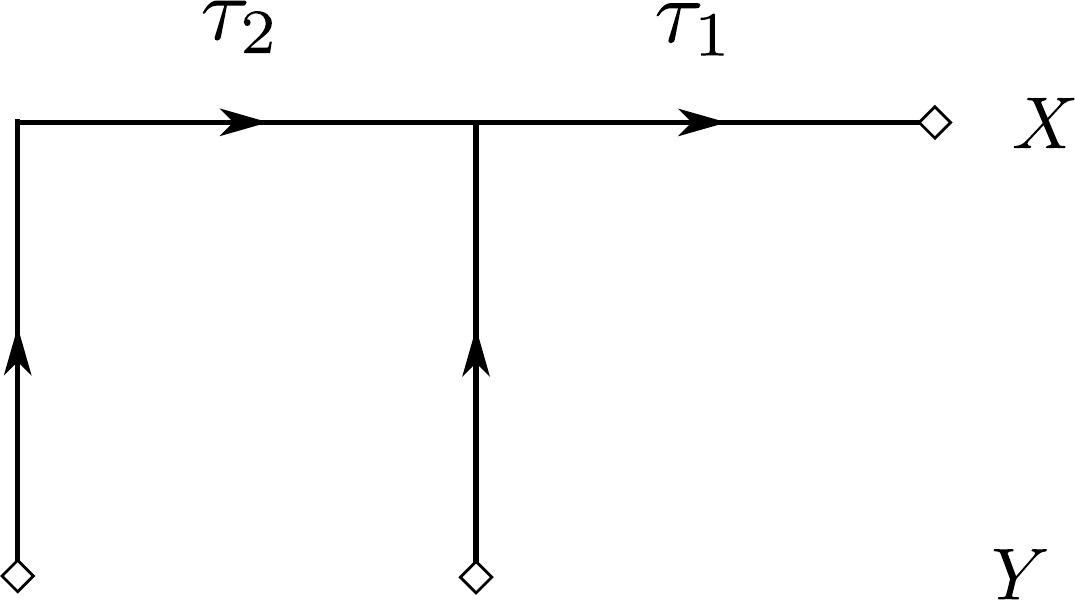}
\caption{Second order response: diagram of the term describing the impact of the fluctuations of the $Y$  variables $G_{A,\Psi,\Psi,a}^{(2)}(\tau_1,\tau_2)$.}
\label{fig:2nd_order1}
\end{figure}

\begin{figure}
\centering
\includegraphics[scale=0.5]{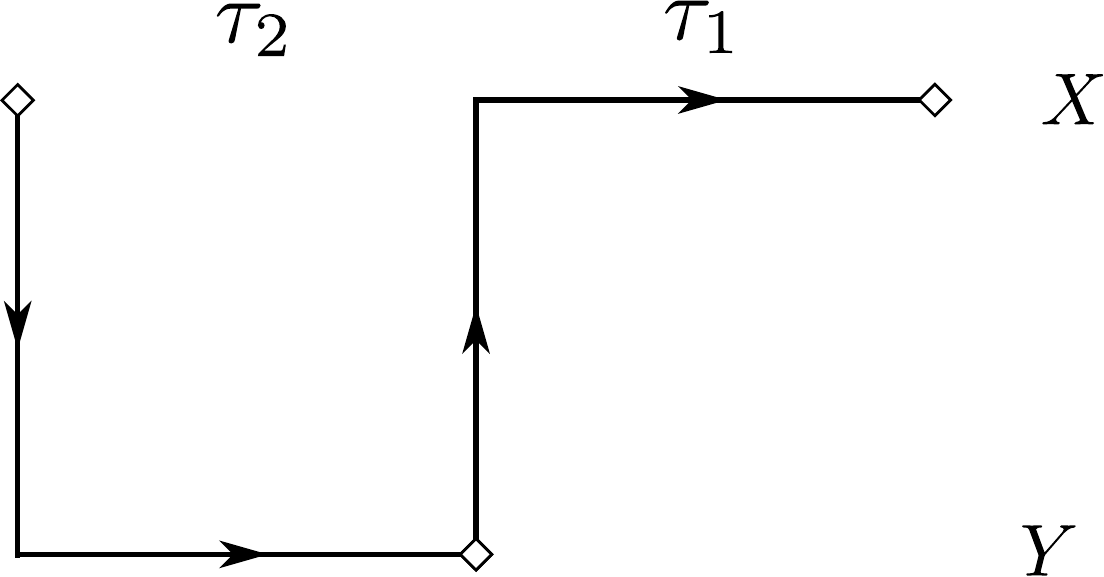}
\caption{Second order response: diagram of the term describing the memory effect of the $X$ variables on themselves mediated by the correlations of the $Y$ variables $G_{A,\Psi,\Psi,b}^{(2)}(\tau_1,\tau_2)$.}
\label{fig:2nd_order2}
\end{figure}

From Eq.~\ref{eq:srb_perturbation} we can derive that at second order the response is a double integral of a response function $G_{A,\Psi}^{(2)}$ by plugging the expressions given in Eqs. \ref{l0} and \ref{l1} into Eq. \ref{drho2}:
\begin{align}
\delta^{(2)}_{\Psi,\Psi} \rho (A) =& \int_0^\infty d s_1 \int_0^\infty d s_2 \; \rho(\mathcal{L}_1 \Pi_0 (s_1)  \mathcal{L}_1 \Pi_0 (s_2) A) = \int_0^\infty d s_1 \int_0^\infty d s_2 G_{A,\Psi,\Psi}^{(2)}(s_1,s_2)\nonumber\\
&=\delta^{(2)}_{\Psi,\Psi,a} \rho (A)+\delta^{(2)}_{\Psi,\Psi,b} \rho (A) =  \int_0^\infty d s_1 \int_0^\infty d s_2 G_{A,\Psi,\Psi,a}^{(2)}(s_1,s_2)+G_{A,\Psi,\Psi,b}^{(2)}(s_1,s_2)  \,,
\label{dpsi2}
\end{align}
where we  separate such a second order contribution into the sum of two terms, $G^{(2)}_{A,\Psi,\Psi,a}$ and $G^{(2)}_{A,\Psi,\Psi,b}$, defined as
\begin{align}
G_{A,\Psi,\Psi,a}^{(2)}(s_1,s_2) = \int \rho_0(dx) \Psi_X(X,Y)\cdot\nabla_X \Pi_0(s_1) \nonumber  (\Psi_X(X,Y)\cdot\nabla_X) (A (f^{s_2}(X)) \,, \\
G_{A,\Psi,\Psi,b}^{(2)}(s_1,s_2) = \int \rho_0(dx)  \Psi_Y(X,Y)\cdot\nabla_Y \Pi_0(s_1)   (\Psi_X(X,Y)\cdot\nabla_X) (A (f^{s_2}(X))  \,.
\label{eq:G_second_order}
\end{align}
The Green function $G_{A,\Psi,\Psi,a}^{(2)}(\tau_1,\tau_2)$ responsible for second contribution $\delta^{(2)}_{\Psi,\Psi,a}$ can be depicted in a diagrammatic mode as in Figure~\ref{fig:2nd_order1}. This term describes the second order effect of the perturbation of the $Y$ dynamics on the statistical properties of the $X$ system, and does not take into account the modifications on the statistical properties of $Y$ due to its coiling with $X$. Instead, Figure ~\ref{fig:2nd_order2} provides a graphical representation of the $G_{A,\Psi,\Psi,b}^{(2)}(\tau_1,\tau_2)$, responsible for the second order contribution  $\delta^{(2)}_{\Psi,\Psi,b}$. In this case, the second order effect comes as a result of the modification to the first order effect shown in Figure ~\ref{fig:1st_order} due to the impact of the $X$ dynamics on the $Y$ systems: note that in this case, as opposed to Figures ~\ref{fig:1st_order}-\ref{fig:2nd_order1}, one arrow points from the $X$ to the $Y$ system.

Our goal is now to provide a second order parametrization of the coupling in terms of the $X$ variables alone, so that  the invariant measure $\hat{\rho}_2$ of the resulting system is such that $\forall A$:
\begin{align}
\rho_{X,Y}(A)=\hat{\rho}_2(A)+O(\Psi^3) \,.\label{o2}
\end{align}
In the following, we show that can be accomplished by the following reduced system:
\begin{align}
\dot{X}(t)&= F_X(X(t)) + M_1(X(t)) + \sum_{i,j} \Psi^\prime_{1,i} (X(t)) \sigma_j (t) + \int_0^\infty d\tau h (\tau,X(t-\tau))\nonumber\\ 
&= F_X(X(t)) + M_1(X(t)) + M_2(X(t))+M_3(X(t)) \,,
\label{eq:second_model}
\end{align}
where the first term on the right hand side given the unperturbed dynamics, the second term is the same given in the first order approximation presented in Eq. \ref{eq:first_model}, the third term is given by a combination of stochastic forcings, and the fourth term describes a memory effect. For simplicity, we indicate all of the perturbative terms as $M_j(X(t))$ with an abuse of notation. We will now provide the explicit expression for the third and fourth terms in Eq. \ref{eq:second_model} by matching the corrections up to $O(\Psi^2)$ to the unperturbed dynamics of the previous with what given in Eqs. \ref{dpsi1} and Eq. \ref{dpsi2}. Obviously, since Eq. \ref{eq:second_model} has a stochastic component, its invariant measure $\hat{\rho}_2$ will be accordingly defined. Please note that in the following, when we refer to first or second order perturbations to the measure $\rho_0$ due to the $M_j$ terms in Eq. \ref{eq:second_model}, the expectation value over the realizations of the stochastic processes are implicitly considered. Note that in \cite{lucarini_stochastic_2011} we have explained how to treat the impact of stochastic perturbations on deterministic dynamical systems along the lines of the Ruelle response theory. 

The first order response in Eq. \ref{eq:second_model} can be written as $\sum_{k}\delta^{(1)}_{M_k}\rho (A) $, where 
\begin{align}
\delta^{(1)}_{M_k}\rho (A)=&\int_0^\infty ds_1 \int \rho_{0,X}(dX) \int_0^\infty d s_1 \;	 G_{A,M_k}^{(1)}(s_1) \nonumber \\
=&\int_0^\infty ds_1 \int \rho_{0,X}(dX) M_k(X) \cdot \nabla_X A(f_X^{s_1}(X))\label{eq:first_M} \,.
\end{align}

The full second order response  can be written as $\sum_{i,j}\delta^{(2)}_{M_i,M_j}\rho (A) $, where 
\begin{align}
\delta^{(2)}_{M_i,M_j}\rho (A)=&\int_0^\infty ds_1 \int_0^\infty ds_2 \int \rho_{0,X}(dX) \int_0^\infty d s_1 \int_0^\infty d s_2 \;	 G_{A,M_i,M_j}^{(2)}(s_1,s_2) \nonumber \\
=&\int_0^\infty ds_1 \int_0^\infty ds_2 \int \rho_{0,X}(dX) M_i(X) \cdot \nabla_X M_j(f_X^{s_2}(X)) \cdot \nabla_{f_X^{s_2}(X)} A(f_X^{s_1+s_2}(X))\label{eq:second_M} \,.
\end{align}
Combining Eqs. \ref{o1} and \ref{o2}, we impose that 
\begin{equation}
\delta^{(1)}_{\Psi}\rho (A) +\delta^{(2)}_{\Psi,\Psi,a}\rho (A)+\delta^{(2)}_{\Psi,\Psi,b}\rho (A) =\sum_{k}\delta^{(1)}_{M_k}\rho (A) +\sum_{i,j}\delta^{(2)}_{M_i,M_j}\rho (A) +O(\Psi^3) \,.
\label{consistency}
\end{equation}
Our procedure will lead to simplifying Eq. \ref{consistency} into the following: 
\begin{equation}
\delta^{(1)}_{\Psi}\rho (A) +\delta^{(2)}_{\Psi,\Psi,a}\rho (A)+\delta^{(2)}_{\Psi,\Psi,b}\rho (A) =\delta^{(1)}_{M_1}\rho (A) +\delta^{(1)}_{M_3}\rho (A)+\delta^{(2)}_{M_1,M_1}\rho (A) +\delta^{(2)}_{M_2,M_2}\rho (A) \,,
\label{consistency2}
\end{equation}
where the terms missing on the right hand side in Eq. \ref{consistency2} with respect to Eq. \ref{consistency} are of order $O(\Psi^3)$ or higher. 

From Eq. \ref{o1}, we have that $\delta^{(1)}_{\Psi}\rho (A)=\delta^{(1)}_{M_1}\rho (A)$. We now consider the next order of perturbation. We first observe that $\delta^{(2)}_{\Psi,\Psi,a}\rho (A)$ can be written as: 
\begin{align}
\delta^{(2)}_{\Psi,\Psi,a}\rho (A) &= \int_0^\infty d s_1 \int_0^{\infty} d s_2  \int \rho_{0}(dx) ((\Psi_X(X,Y)- M(X))\cdot\nabla_X) \Pi_0(s_1)((\Psi_X(X,Y) - M(X))\cdot\nabla_X) (A (f^{s_2}X) \nonumber \\
& + \int_0^\infty d s_1 \int_0^{\infty} d s_2  \int \rho_{0}(dx) (M(X)\cdot\nabla_X) \Pi_0(s_1)((M(X))\cdot\nabla_X) (A (f^{s_2}X) \nonumber \\
& = \int \rho_{0}(dx) (\Psi^\prime_X(X,Y)\cdot\nabla_X) \Pi_0(s_1) (\Psi^\prime_X(X,Y)\cdot\nabla_X) A (f^{s_2}X) +\delta^{(2)}_{M_1,M_1}\rho (A) \nonumber \\
& = \delta^{(2)}_{\Psi,\Psi,f}\rho (A) +\delta^{(2)}_{M_1,M_1}\rho (A) \,,
\label{eq:G_fluct_term}
\end{align}
where $\Psi^\prime_X(X,Y) = \Psi_X(X,Y) - M(X)$ gives the fluctuations of $\Psi$ around its $Y$ averaged value $M$. We now construct the term $M_2(X(t))$ so that, its first order impact $\delta^{(1)}_{M_2}\rho (A)$ vanishes, whereas its second contribution $\delta^{(2)}_{M_2,M_2}\rho (A)$ corresponds exactly to $\delta^{(2)}_{\Psi,\Psi,f}\rho (A)$ introduced in Eq. \ref{eq:G_fluct_term}, which we refer to as the fluctuation term.
%
%
%
\subsubsection*{Fluctuation term}
To analyze the $\delta^{(2)}_{\Psi,\Psi,f}\rho (A)=\int\int ds_1 ds_2 G_{A,\Psi,,\Psi,f}^{(2)}(s_1,s_2)$ term in Eq~\ref{eq:G_fluct_term}, we first consider the case where the coupling fluctuations $\Psi^\prime_X$ are separable in the $X$ and $Y$ variable:
\[\Psi^\prime_X(X,Y)=\Psi^\prime_{X,1}(X) \Psi^\prime_{X,2}(Y) \,. \]
In this case the response function $G^{(2)}_{A,\Psi,\Psi,f}$ can new written as:
\begin{align}
G^{(2)}_{A,\Psi,\Psi,f}(s_1,s_2) = \int \rho_{0,X}(dX) g(s_1) \Psi^\prime_{X,1}(X) \partial_X \Psi^\prime_{X,1}(f_X^{s_1}(X)) \partial_{f^{s_1}_X(X)} (A \circ f^{s_2}_X) (f^{s_1}_X(X)) \,,
\end{align}
where
\begin{equation}
g(s_1)=\langle \Psi^\prime_{X,2}(Y) \Psi^\prime_{X,2}(f_Y^{s_1}(Y)) \rangle_{\rho_{0,Y}} \,. \label{corrnoise}
\end{equation}
We now impose that the second order contributions $\delta^{(2)}_{\Psi,\Psi,f}\rho (A)$ and $\delta^{(2)}_{M_2,M_2}\rho (A)$ are identical, so that:  
\begin{align}
G^{(2)}_{A,\Psi,\Psi,f}(s_1,s_2) &= G^{(2)}_{A,M_2,M_2} (s_1,s_2) \nonumber \\
 &= \int \rho_{0,X} (dX)  M_2 (T - s_1 - s_2, X) \partial_X M_2 ( T - s_1, f^{s_1} (X)) \partial_{f^{s_2}(X)} (A  ( f^{s_1+s_2}(X)) 
\end{align}
for some  time-dependent perturbation $M_2$. This can be achieved by choosing the perturbation $M_2$ to be a multiplicative noise term described as follows:
\begin{align}
M_2(t,X) &= \Psi^\prime_{X,1}(X) \sigma(t) \,, \nonumber \\
\langle \sigma(t_1) \sigma(t_2) \rangle &= g(t_2 - t_1) \,, \nonumber \\
\langle \sigma(t) \rangle &= 0 \,,
\end{align}
where the average in the second equality is an average over realizations of the noise. Taking the average of the model response over the realizations of the noise gives
\begin{align}
\langle G^{(2)}_{A,M_2,M_2}(s_1,s_2) \rangle = G^{(2)}_{A,\Psi,\Psi,f}(s_1,s_2)\Rightarrow\delta^{(2)}_{M_2,M_2}\rho (A)=\delta^{(2)}_{\Psi,\Psi,f}\rho (A) \,.
\end{align}
These results can be easily generalized also to the case where the coupling term $\Psi_X^\prime$ is not separable. In this case,  $\Psi_X^\prime$ can be decomposed in product of Schauder bases of function in $X$ and $Y$:
\[ \Psi_X^\prime(X,Y) = \sum_{i,j} \Psi^\prime_{X,1,i}(X) \Psi^\prime_{X,2,j}(Y) \,. \]
The response function is now given by
\begin{align}
G_{A,\Psi,f}^{(2)\prime} (s_1,s_2) &= \int \rho_{0,X}(dX) \sum_{i,j,k,l} g_{j,l}(s_1) \Psi^\prime_{X,1,i}(X) \partial_X \Psi^\prime_{X,1,k} (f^{s_1}_X(X)) \partial_{f^{s_1}_X} (A \circ f^{s_2}_X) (f^{s_1}_X(X)) \nonumber \\
g_{j,l}(s_1) &= \langle \Psi^\prime_{X,2,j}(Y) \Psi^\prime_{X,2,l} (f^{s_1}_Y(Y)) \,.
\end{align}
The response can in this case be modelled with a linear combination of multiplicative noise
\begin{align}
M_2(X) &= \sum_{i,j} \Psi^\prime_{1,i} (X) \sigma_j (t) \,, \label{eq:subst_fluct} \\
\langle \sigma_j(t_1) \sigma_l(t_2) \rangle &= g_{j,l}(t_2-t_1) \,,\\
\langle \sigma_j(t) \rangle &= 0 \,.
\end{align}
Note also that, since $\langle \sigma_j(t) \rangle=0$ $\forall j$, we have that $\delta^{(1)}_{M_2}\rho (A)=\delta^{(2)}_{M_1,M_2}\rho (A)=\delta^{(2)}_{M_2,M_3}\rho (A)=0$, where we have already anticipated that $M_3$ is not written as a stochastic term. 
At this stage, we have shown that $\delta^{(1)}_{\Psi}\rho (A) +\delta^{(2)}_{\Psi,\Psi,a}\rho (A) =\delta^{(1)}_{M_1}\rho (A) +\delta^{(2)}_{M_1,M_1}\rho (A) +\delta^{(2)}_{M_2,M_2}\rho (A)$. 

\subsubsection*{Memory term}
In order to complete the construction of the parametrization leading to the surrogate dynamics given in Eq. \ref{consistency2}, we need to show that $\delta^{(2)}_{\Psi,\Psi,b}\rho (A)=\delta^{(1)}_{M_3}\rho (A)$, and that the two terms $\delta^{(2)}_{M_1,M_3}\rho (A)$ and $\delta^{(2)}_{M_3,M_3}\rho (A)$ are $O(\Psi^3)$ or higher. The $G_{A,\Psi,\Psi,b}^{(2)}$ term in Eq.~\ref{eq:G_second_order} can be seen as a modification to the first order response on the $X$ system related to the coupling on $Y$ due to the fact that the $Y$ system is modified by the interaction with the $X$ system. After integrating over the $Y$ variables, we can express $\delta^{(2)}_{\Psi,\Psi,b}\rho (A)$ as: 
\begin{equation}
\delta^{(2)}_{\Psi,\Psi,b}\rho (A)=\int_0^\infty ds_1 \int_0^\infty ds_2 \int \rho_{0,X}(dX) \langle \Psi_{Y,i}(X,Y) \partial_{Y,i}  \Psi_{X,j}(f^{s_2}(X),f^{s_2}(Y)) \rangle_{\rho_{0,Y}}\partial_{f^{s_2}(X),j} ( A (f^{s_1+s_2}(X) ) \,.
\end{equation}
This can be expressed as:
\begin{align}
\delta^{(2)}_{\Psi,\Psi,b}\rho (A) =\int_0^\infty d s_1 \int_0^\infty d s_2 \int \rho_{0,X} (dX) h_{j}(s_2,X) \partial_{f^{s_2}(X),j} (A ( f^{s_1+s_2}(X) ) \,, \label{eq:response_2nd_order2}
\end{align}
where we introduce the following integral kernel:
\begin{align}
h_{j} (s,X) = \langle \Psi_{Y,i}(X,Y) \partial_{Y,i}  \Psi_{X,j}(f^{s}(X),f^{s}(Y)) \rangle_{\rho_{0,Y}} \,.
\end{align}
This term describes a memory effect: the evolution of the $Y$ variables between times $T-s_1-s_2$ and $T-s_1$ keeps track (at first order) of the previous state of the $X$ system. Therefore, when we consider the impact of the $Y$ variables on the $X$ system, at second order we get a contribution coming from the history of $X$ itself. Obviously, the faster the decorrelation time of the $Y$ system, the weaker the influence of this term. 

A last step needed at this stage is to provide a response formula for integral forcings, since the formulas derived in Section~\ref{sec:response_theory} are only valid for perturbations without memory. In Appendix~\ref{sec:memory_response} we derive such a generalized expression, which includes as a special case that which has been discussed in Section~\ref{sec:response_theory}. The linear response to a perturbation which can be written as follows:
\begin{equation}
\dot{X} (t) = F_X(X(t))\rightarrow \dot{X} (t) = F_X(X(t))+ \int_0^\infty d\tau \, h (t-\tau,X(t-\tau))
\end{equation}
results to be $\forall$ observables $A$:
\begin{equation}
\delta^{(1)}\rho (A)= \int_0^\infty d s_1 \int_0^\infty d s_2 \int \rho_{0,X} (dX) h_{j}(s_2,X) \partial_{f^{s_2}(X),j} (A ( f^{s_1+s_2}(X) ) \,.
\end{equation}
This indeed implies that by choosing $M_3=\int_0^\infty d\tau h (\tau,X(t-\tau))$ in Eq. \ref{eq:second_model}, we obtain $\delta^{(2)}_{\Psi,\Psi,b}\rho (A)=\delta^{(1)}_{M_3}\rho (A)$. It is important to note that, by construction, even if $\delta^{(1)}_{M_3}\rho (A)$ is a first order perturbation of the expectation value of $A$, it is $O(\Psi^2)$. Therefore, all terms such as $\delta^{(2)}_{M_3,M_3}\rho (A)$, $\delta^{(1)}_{M_1,M_3}\rho (A)$ presented in Eq. \ref{consistency} are $O(\Psi^3)$ or higher, so that Eq. \ref{consistency2}  is indeed correct. Analogously to the case of the noise discussed in the previous section, the term $h_{j} (s,X)$ can be estimated from a trajectory of the system and then used to provide the suitable parametrization. This concludes our derivation and leads to the result anticipated in Eq. \ref{eq:second_model}.
 
\subsection{Independent coupling case}
An especially interesting result can be obtained when considering the not uncommon situation where the coupling is independent of the variable it is affecting, i.e. $\Psi_X(Y)$ depends only on $Y$ and $\Psi_Y(X)$ depends only on $X$. In this case, the reduced dynamical system able to mimic the statistics of the full system for the observables depending on $X$ variables only up to the second order on the coupling strength can be written as:
\begin{align}
\dot{X} (s) = F_X(X(s)) + M_1 + \sigma (s) + \int_0^\infty d\tau \, h (s-\tau,X^\prime(t-\tau)) \,.
\end{align}
The first order term $M_1$ is now a constant drift term and the fluctuation term is an additive noise term with correlations
\begin{align}
\langle \sigma(t_1) \sigma(t_2) \rangle = \langle (\Psi_X(Y) - M_1)(\Psi_X(f^{t_2-t_1}(Y)) - M_1) \rangle_{\rho_{0,Y}} \,.
\end{align}
The integral kernel of the memory term simplifies to
\begin{align}
h(s,X) = \Psi_Y (X) \langle \partial_Y \Psi_X (f^s (Y)) \rangle_{\rho_{0,Y}} \,.
\end{align}
where the role of the impact of the memory effect of the $Y$ variables on the memory of the $X$ system is especially clear.
 
\section{Comparison with other variables reduction methods}\label{comparison}
The problem of variables reduction  is a classic topic in dynamical systems theory and statistical mechanics. We will briefly discuss the relation between these methods and what presented here. A well-developed approach is the so called averaging method \cite{arnold_hasselmanns_2001, kifer_recent_2004}. This method deals with coupled systems of the form
\begin{align}
\dot{X} &= f(X,Y) \,, \\
\dot{Y} &= \frac{1}{\epsilon} g(X,Y) \,.
\end{align}
In the limit where the time-scale difference between the two goes to infinity, by taking $\epsilon \rightarrow \infty$, the trajectories of the original system converge on finite time scales to the trajectories of an averaged version of the $X$ system, namely
\begin{align}
\dot{X^\prime}=\rho_{Y|X} \left( f (X^\prime,Y) \right) \,,
\end{align}
where $Y$ is average over the measure $\rho_{Y|X}$, which is the SRB measure corresponding to the dynamics of $Y$ for fixed $X$:
\begin{align}
\dot{Y} &= g_X (Y) \,, \\
g_X(Y) &= g(X,Y) \,.
\end{align}
The fluctuations around this average trajectory can be approximated by an Ornstein-Uhlenbeck process. One should note that while we have discussed here long term averages in the form of SRB measures, the averaging method describes convergence on finite time scales. Another point of difference is the weak coupling assumption we have made here. This assumption is not made in the averaging method, where the limit of an infinitely fast $Y$ process is taken instead. Such a limit basically takes care of getting rid of the higher order effects discussed in this paper.

In~\cite{abramov_suppression_2011} the averaging method is used to derive a simplified reduced model. The $X$ dependent measure $\rho_{Y|X}$ is first replaced by the measure $\rho_{Y|X^*}$ for a fixed value $X^*$ for $X$. To improve the statistics of the reduced model a linear term in $X$ is added:
\begin{align}
\dot{X}^\prime = \rho_{Y|X} \left( f (X^\prime,Y) \right) + (X-X^*) \delta \rho_{Y|X^*} \left( f (X^\prime,Y) \right) \,,
\end{align}
where $\delta \rho_{Y|X^*}$ is the response of the measure $\rho_{Y|X^*}$ to a change in the value of the fixed $X$. Such a feedback mechanism, where a change in $X$ affects the $X$ dynamics through the presence of $Y$, is also present in the memory effect we find here. However, as the limit of infinite time-scale differences is taken in the averaging method, there the feedback happens instantly from the point of view of the slow process and no memory is present.

Projector operator techniques  have found much use in the statistical mechanics for deriving Fokker-Planck equations. A celebrated result in the Mori-Zwanzig equation, which consists of an integro-differential equation, containing an integral term describing the memory effect due to projected out variables~\cite{zwanzig_nonequilibrium_2001}.
Consider the setting of a dynamical system with Liouville operator $L$ where we have a projector operator $P$ projecting out irrelevant parameters. Every observable $A$ can the be then be written as a two-vector containing $A_1=PA$ and $A_2=QA=(1-P)A$. The Liouville equation for this vector is then given by
\begin{align}
\frac{d}{dt} \begin{pmatrix}
A_1 \\
A_2
\end{pmatrix}
=
\begin{pmatrix}
L_{11} & L_{12} \\
L_{21} & L_{22}
\end{pmatrix} 
\begin{pmatrix}
A_1 \\
A_2
\end{pmatrix} \,,
\end{align}
where $L_{11}=PLP$, $L_{12}=PLQ$, $L_{21}=QLP$ and $L_{22}=QLQ$. By using Duhamel's formula for the equation for $A_2$ a solution for $A_2$ can be found where the term with $L_{21}A_1$ is seen as a perturbation to the free evolution of $A_2$ under $L_{22}$. By plugging this solution back into the differential equation for $A_1$, the following equation of motion for $A_1$ is obtained
\begin{align}
\frac{d}{dt} A_1(t) = L_{11} A_1(t) + L_{12} \int_0^t ds \exp(L_{22}(t-s)) L_{21} A_1(s) + L_{12} \exp(L_{22}t)A_2 (0) \,.
\end{align}
Again there is a feedback mechanism through the presence of the irrelevant dynamics. It has to be noted that the Mori-Zwanzig equation is not an approximation, but an equivalent formulation of the original problem. This approach is concerned with describing the instantaneous evolution of probability measures, describing the evolution of the current state, rather than the long term averages, as discussed here. Our strategy may provide a way to compute, at least approximately the projector operators that in the Mori-Zwanzig theory are defined only formally.

\section{Conclusions}\label{sec:conclusions}

In this paper we have adopted the Ruelle response theory to study the effect of a weak coupling between two dynamical systems. We have derived explicit formulas to compute the changes in the expectation value of a general observable of one of the two systems (we call this system $X$, and refer to the other system as $Y$) up to the second order in the coupling strength. We have also presented a simple diagrammatic representation of such terms, which is suggestive of the possibility of reframing the Ruelle theory in graphical terms along the lines of the Feynman representation of quantum electrodynamics. 

These results allow us to define a surrogate perturbed dynamics for the system $X$ such up to second order its statistical properties are the same as those of the coupled system when a general observable of $X$ alone is considered. In other terms, we have attempted to derive a parametrization of the coupling. We have found that the invariant measure for the coupled system restricted to the $X$ system can be approximated by the invariant measure of a reduced system with three forcing terms $M_i$:
\begin{align}
\frac{d}{dt} X(t) = F_X(X(t)) + M_1(X(t)) + M_2(X(t))+M_3(X(t)),
\end{align}
where $F_X(X(t))$ gives the unperturbed flow, $M_1$ is a deterministic correction, which fully describes the change in the invariant measure up to first order, $M_2$ is a stochastic forcing expressed as a sum of multiplicative noise terms with non-trivial correlation properties, and $M_3$ is an integral expression which describes a memory term. The explicit expressions for the $M_j$ terms are given in Eq. \ref{eq:second_model}. The $M_1$ term basically describes a mean field theory, where the effective $X$ dynamics is modified by adding the effect of the $Y$ averaged dynamics. The $M_2$ term, instead, takes care of representing the effect of the fluctuations of the $Y$ systems. Finally, the $M_3$ term describes the non-markovian effect due to the $X$ system impacting the state of the $Y$ state at a given time in the past, which in turn perturbs the $X$ system at a later time.  Note that, if the coupling is independent of the variable it is affecting, so that the perturbation on $X$ ($Y$) depends only on $Y$ ($X$) variables, the parametrization becomes especially attractive, as $M_1$ reduces to a constant, $M_2$ to an additive noise term and $M_3$ has a greatly simplified integral kernel. 

It is important to note that in our derivation no assumptions have been made on the dynamics of $X$ and $Y$ systems in terms of differing time scales and/or of decorrelation time, our only hypothesis being the assumption that the coupling is weak in some sense: there is - a priori - symmetry between the $X$ and $Y$ systems, we just select for which one we want to derive an effective dynamics. Actually, the characteristics of the unperturbed dynamics of the two systems are expected to play an important role in the relative importance of the fluctuation and memory terms. In a system that is highly mixing, correlations in time disappear fast and the fluctuation term will act as white noise. The memory term on the other hand will quickly drop off to zero in this case. If the $Y$ system decorrelates slowly, there will be a more pronounced memory effect.

It is also important to note that our method does not aim at describing within a certain accuracy and for a certain time horizon the individual orbits of the system of interest, but rather focuses on providing an approximate method for reconstructing its statistical properties. Therefore, the method discussed in this paper has points of contact and differences with variable reduction techniques previously proposed in the literature such as the averaging method, which assumes a very large time scale separation but no limit on the strength of the coupling, and the projection operator method which instead is more formal and has less straightforward practical applicability.  

Obviously, if the system consists of more than two levels it can be reduced to only one by applying repeatedly to the approach described in this paper, so that our method has a good degree of generality when variables reduction is sought.

These findings will be of particular interest in the scientific fields dealing with multilevel systems and in particular, in the context of climate modelling. First of all we have derived an explicit strategy to substitute unresolved processes that is robust to changes in parameters or scales. Furthermore are the parameters (such as the mean coupling, the noise correlations and the memory kernel) explicitly formulated in terms of statistical properties only of the unresolved processes. By comparing the fluctuation and memory terms we see that different strategies will be applicable whether the unresolved process is fast or slow. Hence, this is relevant when modelling, e.g., the impact of the atmosphere on the oceans versus the other way around.

As future research we will be implementing the proposed modelling strategies in low-dimensional systems, such as the Lorenz '96 system~\cite{lorenz_predictability:_1996}. Hopefully this can lead to applications on more realistic models. Of particular interest will be how well this derivation based on weak couplings will apply to couplings in the climate system and whether the interplay between the fluctuations and memory can be observed.

\section*{\small{Acknowledgements}}
The authors acknowledge Matteo Colangeli for fruitful discussions.

\newpage
\appendix
\section{Response to perturbations with memory}
\label{sec:memory_response}
As in~\cite{lucarini_relevance_2011}, we derive the response of a discrete dynamical system
\[ x_{t+1} = f(x_t) \,.\]
Consider a perturbation $\chi$ that has a memory of the variable at $l$ time steps in the past:
\[ \tilde{x}_{t+1} = f(\tilde{x}_t) + \chi_l(\tilde{x}_{t-l}) \,. \]

We can expand evolution of $\tilde{x}$ around the evolution of $x$. By induction, one can show that
\begin{align}
\tilde{x}_{t+k} = f^k(\tilde{x}_t) + \sum_{j=1}^k Df^{k-j} (f^j(\tilde{x}_t)) \chi_l(\tilde{x}_{t-l+j-1}) + O(\chi^2) \,,
\end{align}
or on an observable $A$ that
\begin{align}
A(\tilde{x}_{t+k}) = A(f^k(\tilde{x}_t)) + \sum_{j=0}^k D(A \circ f^{k-j}) (f^j(\tilde{x}_t)) \chi_l(\tilde{x}_{t-l+j-1}) + O(\chi^2) \,.
\end{align}
From this we derive the following linear response
\begin{align}
\tilde{\rho} (A) &= \lim_{T \rightarrow \infty} \frac{1}{T} \sum_{k=1}^T A(\tilde{x}_{t+k})) = \rho(A) + \frac{1}{T}  \sum_{k=1}^T \sum_{j=1}^k \chi_l(f^{j-1}(\tilde{x}_{t-l})) D(A \circ f^{k-j}) (f^{j+l}(\tilde{x}_{t-l})) + O(\chi^2) \\
&=  \rho(A) + \int \rho(dx)  \sum_{i \geq 0} \chi_l(x) D(A \circ f^{i}) (f^{l+1}(x) ) + O(\chi^2) \,. \label{eq:mem_response_l}
\end{align}

Let's consider the case where the memory dependence is given by a sum of memory terms at different times in the past:
\begin{align}
 \tilde{x}_{t+1} = f(\tilde{x}_t) + \sum_{l=0}^{\infty} \chi_l \left(\tilde{x}_{t-l} \right) \,.
\end{align}
The response can be derived from Eq.~\ref{eq:mem_response_l} by linearity:
\begin{align}
\delta \rho (A) = \int \rho(dx) \sum_{l=0}^\infty \sum_{i \geq 0} \chi_l(x) D(A \circ f^i) (f^{l+1}(x)) + O(\chi^2) \,.
\end{align}
In the continuous case, for an equation of the form
\begin{align}
\frac{d}{dt} \tilde{x}_t = F(\tilde{x}_t) + \int_0^\infty \chi_\tau (\tilde{x}_{t-\tau})
\end{align}
this response becomes
\begin{align}
\delta \rho_T = \int \rho(dx) \int_0^\infty ds \int_0^\infty d\tau \chi_{\tau}(x) \partial_{f^{\tau}(x)} (A \circ f^{s}) (f^{\tau}(x)) + O(\chi^2) \,. \label{eq:response_memory}
\end{align}


\newpage

\bibliographystyle{plain}
\bibliography{stochastic_modelling}

\newpage

\end{document}